# Marangoni effect and cell spreading

Ivana Pajic-Lijakovic*, Milan Milivojevic

Faculty of Technology and Metallurgy, Department of Chemical Engineering, University of Belgrade, Serbia

Correspondence to Ivana Pajic-Lijakovic, email: iva@tmf.bg.ac.rs

**Abstract**

Cells are very sensitive to the shear stress (SS). However, undesirable SS is generated during physiological process such as collective cell migration (CCM) and influences the biological processes such as morphogenesis, wound healing and cancer invasion. Despite extensive research devoted to study the SS generation caused by CCM, we still do not fully understand the main cause of SS appearance. An attempt is made here to offer some answers to these questions by considering the rearrangement of cell monolayers. The SS generation represents a consequence of natural and forced convection. While forced convection is dependent on cell speed, the natural convection is induced by the gradient of tissue surface tension. The phenomenon is known as the Marangoni effect. The gradient of tissue surface tension induces directed cell spreading from the regions of lower tissue surface tension to the regions of higher tissue surface tension and leads to the cell sorting. This directed cell migration is described by the Marangoni flux. The phenomenon has been recognized during the rearrangement of (1) epithelial cell monolayers and (2) mixed cell monolayers made by epithelial and mesenchymal cells. However, proper explanation of this phenomenon has not provided. The consequence of the Marangoni effect is an intensive spreading of cancer cells through an epithelium. In this work, a review of existing literature about SS generation caused by CCM is given along with the assortment of published experimental findings, in order to invite experimentalists to test given theoretical considerations in multicellular systems.

1. **Introduction**

Tissue rearrangement during morphogenesis, wound healing, and cancer invasion depends on the coordinated migration of cellular clusters (Clark and Vignjevic, 2015; Barriga et al., 2018; Barriga and Mayor, 2019; Pajic-Lijakovic and Milivojevic, 2019a;2020a). The success of these clusters in reaching their target tissues relies on their ability to migrate in a synchronized way (Barriga and Mayor, 2019). The tissue rearrangement depends on behaviors of cell movement such as: coordination, cooperation, and supracellularity (Shellard and Mayor, 2019). Cell coordination correlates with the directional cell movement, whereas cluster cooperation is determined by the characteristics of cell-cell adhesion interactions. The supracellularity is a term used to describe a group of cells with a high level of intracellular organization (Shellard and Mayor, 2019). The directional cell movement normally relies on a variety of external signals, such as chemical, mechanical, or electrical, which instruct cells in which direction to move (Shellard and Mayor, 2020). Consequently, the directional cell movement has been described by various fluxes such as: chemotaxis, electrotaxis, durotaxis expressed as function of the nutrient concentration gradient, gradient of electrostatic field, stiffness gradient of extracellular matrix, respectively (Murray et al., 1988). Besides these fluxes, the Marangoni flux (which represents a consequence of the gradient of tissue surface tension) also contributes to the directional cell movement. The Marangoni effect exists in various soft matter systems, but has not been discussed in the context of CCM. The surface tension gradient can be induced by changing the temperature or distribution of constituents within the soft matter system (Karbalaei et al., 2016). The surface tension gradient guides the system shear flow from the regions of lower surface tension to the regions of higher surface tension (Karbalaei et al., 2016). This shear flow induces a generation of corresponding SS as the consequence of natural convection. The tissue surface tension gradient also can be established within multicellular systems (Devanny et al., 2021).

The tissue surface tension gradient can be established between (1) contractile (migrating) epithelial cell clusters and surrounding non-contractile (resting) epithelium, (2) migrating mesenchymal cell clusters and surrounding migrating or resting epithelium. It is in accordance with the fact that migrating clusters of healthy epithelial cells have higher tissue surface tension then the clusters of resting (non-contractile) cells (Devanny et al., 2021). In the contrast to healthy epithelial cells, the migrating clusters of mesenchymal cells have lower tissue surface tension in comparison with the clusters of resting mesenchymal cells (Devanny et al., 2021). Migrating mesenchymal cells also have a lower tissue surface tension then contractile and non-contractile epitheliums (Devanny et al., 2021). Some authors considered CCM of mixed cell monolayers consisted of epithelial and mesenchymal cells and pointed out that presence of epithelium stimulate movement of mesenchymal cells (Lee et al., 2012; Heine et al., 2021; Wu et al., 2021). Mesenchymal cells are capable of establishing higher level of coordination and higher speeds within a contractile epithelium (Heine et al., 2021; Wu et al., 2021).

While the gradient of tissue surface tension contributes to the cell SS, the surface tension itself contribute to the cell normal stress. The relationship between the tissue surface tension and the cell normal stress is expressed based on the Young-Laplace equation (Marmottant et al., 2009; Pajic-Lijakovic and Milivojevic, 2019b). The cell SS generation during CCM occurs via natural and forced convection. The natural convection is caused by inhomogeneous distribution of the tissue surface tension, while the forced convection is induced by movement of migrating cell clusters and depends on the speed of clusters. Tambe et al. (2013) considered stress generation within collective migrated Madin-Darby canine kidney type II (MDCK) cell monolayers. They measured SS in the range of *100-150 Pa* by using the 2D monolayer stress microscopy (MSM). Patel et al. (2020) considered CCM of endothelial monolayers by using MSM and emphasized that shear and normal inter-cellular tractions are the same order of magnitude. The proposed stress measuring procedure applied by Tambe et al. (2013) and Patel et al. (2020) accounts for the estimation of inter-cellular stress distribution computed for a measured traction field. However, Green et al. (2020) pointed out that a direct validation of the 2D stresses predicted by a linear passive MSM model is presently not possible. Measuring of the stress distribution especially in 3D is a difficult task. Several recent studies introduced microbead/droplet-based stress sensors of well-controlled mechanical properties to 3D cellular systems in order to overcome the complexities associated with the development of 3D traction force microscopy (Zhang et al., 2019). Dolega et al. (2017) used elastic microbead sensors to measure isotropic compressive stress caused by tumor growth within the matrix. Incompressible micro-droplet sensors are able to measure anisotropic normal stresses (Campas et al., 2013). Zhu et al. (2020) recently developed a magnetic device for measuring 3D stiffness distribution. Development of suitable devices represents a prerequisite for improving our knowledge about this complex phenomenon. Consequently, we can only estimate the value of generated SS. The SS of *100-150 Pa* is too large, while the value of several tens of Pa could be expectable.

Undesirable SS generation during morphogenesis, wound healing, regeneration, and cancer invasion represents a frequent problem for regenerative medicine and can induce various diseases (Flitney et al., 2009; Molladavoodi et al., 2017; Pitenis et al., 2018; Delon et al., 2019). Cells response depends on the order of magnitude of SS and the rate of its change. Even a very low SS, $< 1\ Pa$ can provoke various molecular mechanisms which lead to: cell shape changes, gene expression, cytoskeleton softening, remodeling of cell-cell and cell-matrix adhesion contacts, and can induce epithelial-to-mesenchymal transition (EMT) (Flitney et al., 2009; Delon et al., 2019; Pitenis and Sawyer, 2020). These changes at a subcellular level influence the cell migration and proliferation (Molladavoodi et al., 2017). Higher shear stress $> 10\ Pa$ can induce cell inflammation and even cell death (Rahman et al., 2018; Pitenis and Sawyer, 2020). Pitenis et al. (2018) considered the response of human corneal epithelial (hTCEpi) cell monolayers under SS generated at the biointerface with the hydrogel made by 7.5 wt% polyacrylamide and 0.3 wt% bisacrylamide. They revealed that the SS of *60 Pa* is sufficient to induce an inflammation of cells and fibrosis during 5.5 h.

The aim of this contribution is to discuss (1) the possible causes of cell sensitivity to SS from the standpoint of mechanics and rheology and (2) the origin of SS generation caused by 2D CCM of healthy epithelial cells and mesenchymal cells as a consequence of natural and forced convection.

2. **Cell sensitivity to shear stress**

Cell response to lower SS $\leq 1\ Pa$ can induce a remodeling of the cell cytoskeleton, as well as, cell-cell and cell-matrix adhesion contacts. Various biochemical processes influence the cytoskeleton remodeling such as: phosphorylation and EMT (Flitney et al., 2009). Delon et al. (2019) considered the response of Caco2 cell monolayer under flow SS. They reported that the corresponding monolayer thickness is 10 µm under flow SS of $\sim 2x10^{-3}\ Pa$. An increase in flow SS to $\sim 3x10^{-3}\ Pa$ induces an increase in the monolayer thickness to *28 µm*. Molladavoodi et al. (2017) reported that exposure of corneal epithelial cells to flow SS of 0.4 and 0.8 Pa, induces more prominent, organized and elongated filamentous actin. Higher rates of migration and proliferation were observed under lower flow SS of *0.4 Pa* within 24 h. Larger flow SS of *0.8 Pa* is capable of inducing cell damage and intensive cell apoptosis. Oscillatory flow SS with a peak of *4 Pa* leads to an increase in the gene expression of E cadherin and tight junction protein 1B in comparison with constant flow SS of *1.4 Pa*. The mRNA expression of E-cadherin and occludin were increased under oscillatory medium flow which reinforced adherens junctions (AJs) and tight junctions (TJs). Desmoplakin and occludin proteins were up-regulated under oscillatory SS. Stress fiber formation was not aligned to flow direction. MUC1, -4, and -16 protein were localized under all culture conditions, a regulation on mRNA expression was not detectable (Molladavoodi et al., 2017). However, higher SS of several tens of Pa can induce inflammation of cells, the membrane local disintegration, and even cell programmable death. However, despite extensive research devoted to study mechanical and biochemical responses of cells under SS, we still do not fully understand the main cause of cells sensitivity to the SS. The cause could be connected with the rheological behavior of semi-flexible filaments such as microfilaments and microtubules rather than intermediate filaments which are more flexible. The SS induces disordering in the state of the cytoskeletal filaments. Some filaments are stretched while the neighbors are compressed at the same time. The response of single-filaments depends on their current orientation relative to the direction of external SS. Stretched semi flexible filaments exert larger force than the compressed ones under the same absolute deformation (Broedersz and MacKintosh, 2014). The phenomenon is accompanied with the nonlinear force change during stretching or compression of single semi-flexible filaments, described by the worm-like chain model (Yamakawa, 1971). It is presented schematically on **Figure 1**.

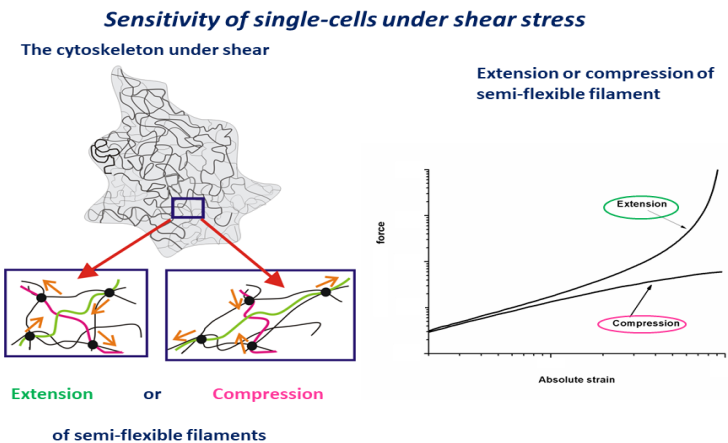

**Figure 1**. Semi-flexible filaments of the cytoskeleton under stretching and compression.

Stress fibers (SFs), a one of the main contributors to cell mechanics, have been treated as the worm-like chains (Bathe et al., 2008). The SF represents a contractile bundle of 10-30 actin filaments (Deguchi et al., 2006). Various types of SFs such as: (1) ventral SFs, (2) dorsal SFs, (3) perinuclear actin cap, and (3) actin arcs represent the constituents of SFs network (Kassianidou and Kumar, 2015). Deguchi et al. (2006) considered tensile properties of single actin-myosin SFs, isolated from cultured vascular smooth muscle cells. They reported that the amount of force required for stretching the single SFs from the zero-stress length back to the original length was approximately 10 nN. This stretching is accompanied with the entropic effects. Larger forces induce entalpic effects which cause the non-linear rheological response of SFs. The force of several tens of nN corresponds to an order of magnitude of cell traction force (Emon et al., 2021). In contrast, the rheological response of compressed SFs is primarily linear due to entropic effects (Broedersz and MacKintosh, 2014). If the shear force of 10 nN acts on the surface $l_c X l_c$ (where $l_c$ is the size of single cell equal to $l_c \approx 10\ \mu m$), the corresponding SS is equal to $30\ Pa$. This value of SS is enough to partially disintegrate the cytoskeleton. The SS itself induces the generation of normal stress within the cytoskeleton. Janmey et al. (2006) pointed to the negative first normal stress difference, as the characteristic of a semi-flexible filament network response, under the Couette shear flow. Consequently, the structural ordering of semi-flexible filaments under SS leads to an inhomogeneous accumulation of strain energy which can induce partial disintegration of the cytoskeleton.

Cell sensitivity to SS and resistance to compressive stress can be roughly quantified by Young's modulus and shear modulus of single cells, respectively. Larger value of the corresponding modulus (shear or Young's modulus) points to a more resistant structure. Eldridge et al. (2019) measured the apparent Young's modulus of MCF7 and BT474 breast cancer cells by using the atomic force microscopy and shear modulus by using the quantitative phase imaging. The cell shear modulus was measured under flow SS of *0.35, 0.7* and *1.4 Pa*. They revealed that the Young's modulus was approximately three times higher than the shear modulus which can be valid only for the isotropic elasticity and Poisson's ratio of 0.5. However, cells are anisotropic, inhomogeneous, and viscoelastic even at time scale of milliseconds to seconds (Pajic-Lijakovic and Milivojevic, 2015). Consequently, it is expectable that the ratio between shear modulus and Young's modulus can be lower than that proposed by Eldridge et al. (2019).

### 3. Shear stress generation caused by CCM

The origin of SS generation caused by CCM is discussed on the model systems such as the rearrangement of cell monolayers. Cell monolayers are inhomogeneous and have been treated as an ensemble of microdomains (Pajic-Lijakovic and Milivojevic, 2021a). Various types of cell domains were distinguished as (1) domains of free or weakly connected active (contractile) cells which undergo highly coordinated single-cell migration i.e. the multicellular streams (Friedl and Alexander, 2011), (2) domains of strongly connected migrating clusters (Pajic-Lijakovic and Milivojevic, 2020b), and (3) stagnant zones formed by the collision of velocity fronts (Nnetu et al., 2012). The rearrangement of multicellular systems has been considered as the distribution various domains (Pajic-Lijakovic and Milivojevic, 2021a).

Cell movement within a domain as well as the effects at the biointerface between domains can induce natural and forced convection. Total SS can be expressed as the sum of these two contributions as:

$$\vec{n} \cdot \tilde{\sigma}_{cRS} \cdot \vec{t} = \vec{\nabla}\gamma \cdot \vec{t} + \vec{n} \cdot \tilde{\sigma}_{cRS}^{F} \cdot \vec{t} \tag{1}$$

where $\tilde{\sigma}_{cRS}$ is the total cell residual SS, $\tilde{\sigma}_{cRS}^{F}$ is the SS generated as a consequence of the forced convection, $\gamma$ is the tissue surface tension, $\vec{t}$ is the tangent vector of the surface, and $\vec{n}$ is the normal vector of the surface. The first term on the right hand side accounts for the effect of natural convection while the second term represents the consequence of forced convection. The SS generation by natural convection will be discussed in the context of the tissue surface tension gradient $\vec{\nabla}\gamma$. The SS generation by forced convection represents the difference between total SS and the SS generated by natural convection. The total SS generation will be discussed based on two constitutive models frequently used for describing CCM of multicellular systems, i.e. the Maxwell model and the Zener model (Pajic-Lijakovic, 2021).

### 3.1 Shear stress generation by natural convection

The SS generation by natural convection represents a consequence of an inhomogeneous distribution of the tissue surface tension. The tissue surface tension depends on: (1) cell type, (2) the type of cell-cell and cell-ECM adhesion contacts, (3) cell active (contractile) or passive (resting) state, and (4) configuration of migrating cells. It is influenced by various biological processes such as: (1) contact inhibition of locomotion (CIL) (Lin et al. 2018), (2) cell polarization alignment (AL), and (3) EMT (Alert and Trepat, 2020). An increase in cell packing density $n_c(r,\tau)$ intensifies the CIL which results in decrease in the tissue surface tension (Alert and Trepat, 2020). The EMT is a cellular process during which epithelial cells acquire mesenchymal phenotypes (Yang et al., 2020). They pointed out to the main characteristics of EMT: (1) cytoskeleton remodeling, (2) loss of apical–basal cell polarity, (3) cell–cell adhesion weakening, (4) cell–matrix adhesion remodeling, (5) cell individualization, (6) establishment of the front–back cell polarity, (7) acquisition of cell motility, and (8) basement membrane invasion. Epithelial cells are stiffer than the mesenchymal cells. The cell stiffness can be quantified by the Young's modulus. Lekka et al. (1999) reported that the Young's modulus of bladder cancer cells is $E = 14.0 \pm 2.2 \ kPa$, while for non-malignant ones the modulus is $E = 28.5 \pm 3.9 \ kPa$. Beside the mechanical state of single cells, the state of AJs also contributes to the tissue surface tension. Weakening of AJs, characteristic for the mesenchymal cells can be caused by decreasing in the concentration of E-cadherin and in some cases, by increasing in the concentration of N-cadherin (Barriga and Mayor, 2019). The SS of only *0.14 Pa* is sufficient to induce the EMT in Hep-2 cells (Liu et al., 2016). The SS of *0.3 Pa* causes the EMT in epithelial ovarian cancer (Rizvi et al., 2013). Consequently, EMT induces a decrease in the tissue surface tension.

An inhomogeneous distribution of the tissue surface tension is occurred within mixed multicellular systems. The mixing of cells within cellular tissues is a crucial property for diverse biological processes, ranging from morphogenesis, immune action, to tumor metastasis. Devanny et al. (2021) considered the rearrangement and compaction of various breast cell aggregates such as: healthy epithelial MCF-10A cells and various mesenchymal cell lines, i.e. MDA-MB-468, along with MDA-MB-231 and MDA-MB-436 cells. They suggested that the contractility plays a fundamentally different role in the cell lines in which compaction is driven primarily by integrins (MDA-MB-468, along with MDA-MB-231 and MDA-MB-436 cells) vs. by cadherins (MCF-10A cells). Cell contractility influences the state of cell-cell and cell-ECM as well as their crosstalk. The contractility of healthy epithelial MCF-10A cells induces reinforcement of E-cadherin mediated cell-cell adhesion contacts and, on that base, leads to an increase in the tissue surface tension. Otherwise, enhanced contractility of cancer cells causes the cell-cell repulsion and leads to a decrease in the tissue surface tension.

Consequently, several types of multicellular systems will be considered in the context of inhomogeneous distribution of the tissue surface tension. These systems were presented in Table 1.

**Table 1**. Multicellular systems which establish the surface tension gradient

| System | Migrating cell cluster | Surrounding cells | Tissue surface tension |
|---|---|---|---|
| 1 | Epithelial cluster | Epithelium in the non-contractile state | $\gamma_{EC}^{c} > \gamma_{EC}^{Nc}$ |
| 2 | Epithelial cell cluster | Mesenchymal cells in the non-contractile state | $\gamma_{EC}^{c} > \gamma_{MC}^{Nc}$ |
| 3 | Epithelial cell cluster | Mesenchymal cells in the contractile state | $\gamma_{EC}^{c} > \gamma_{MC}^{c}$ |
| 4 | Mesenchymal cell cluster | Mesenchymal cells in the non-contractile state | $\gamma_{MC}^{c} < \gamma_{MC}^{Nc}$ |
| 5 | Mesenchymal cell cluster | Epithelium in the non-contractile state | $\gamma_{MC}^{c} < \gamma_{EC}^{Nc}$ |
| 6 | Mesenchymal cell cluster | Epithelium in the contractile state | $\gamma_{MC}^{c} < \gamma_{EC}^{c}$ |

where $\gamma_{EC}^{c}$ is the tissue surface tension of contractile epithelium, $\gamma_{EC}^{Nc}$ is the tissue surface tension of non-contractile epithelium, $\gamma_{MC}^{c}$ is the tissue surface tension of contractile mesenchymal cells, and $\gamma_{MC}^{Nc}$ is the tissue surface tension of non-contractile mesenchymal cells

Migrating clusters of healthy epithelial cells have higher tissue surface tension then the clusters of resting (non-contractile) cells (Devanny et al., 2021). In the contrast to healthy epithelial cells, the migrating clusters of mesenchymal cells have a lower tissue surface tension in comparison with the clusters of resting mesenchymal cells (Devanny et al., 2021). The surface tension gradient is established between: migrating cell clusters surrounding cells.

The gradient of surface tension can be expressed as $\frac{\Delta\gamma}{\Delta L}$ (where $\Delta\gamma$ is the difference in the tissue surface tension between migrating cell clusters and surrounding cells and $\Delta L$ is the characteristic length. Mombash et al. (2005) considered a change in the tissue surface tension during the rounding of 3D chicken embryonic neural retina aggregates. They reported that the tissue surface tension increases from $1.6 \pm 0.6 \frac{mN}{m}$ to $4.0 \pm 1.0 \frac{mN}{m}$ within 9 days. Beysens et al. (2000) found that the values of the average tissue surface tension for five chicken embryonic tissues varied from $1.6\ \frac{mN}{m}$ (neural retina) to $20\ \frac{mN}{m}$ (limb bud). The tissue surface tension of embryonic carcinoma F9 cell line is $4.74 \pm 0.28\ \frac{mN}{m}$ (Stirbat et al., 2013). The gradient of tissue surface tension has not been measured yet but can be estimated in order to provide preliminary value of the SS generated by the natural convection. For supposing $\Delta \gamma \approx 2\ \frac{mN}{m}$ (which corresponds to the experimental data by Mombash et al. (2005) and $\Delta L \approx 100\ \mu m$ (which is an order of magnitude higher than the size of single cell), the calculated gradient of surface tension as well as the SS part generated by the natural convection can be estimated as $\sim 20\ Pa$.

Gradient of the tissue surface tension directs cell spreading from the regions of lower tissue surface tension to the regions of higher tissue surface tension. This directional CCM can be expressed by the flux:

$$\vec{J}_M = k_M n_c(r,\tau) \vec{\nabla} \gamma \qquad (2)$$

where $\vec{J}_M$ is the Marangoni flux, $k_M$ is the model parameter which represents a measure of cell mobility, $n_c(r,\tau)$ is the cell packing density. The Marangoni flux together with convective, conductive, haptotaxis, galvanotaxis, chemotaxis fluxes et so on contributes to the change in cell packing density. The overall mass balance was formulated by Murray et al. (1988) as:

$$\frac{\partial n_c(r,\tau)}{\partial \tau} = -\vec{\nabla} \cdot \vec{J} \qquad (3)$$

where $\vec{J}$ is the flux of cells equal to $\vec{J} = \vec{J}_{conv} + \vec{J}_{cond} + \sum_i \vec{J}_i$, such that $\vec{J}_{conv} = n\vec{v}_c$ is the convective flux, $\vec{v}_c$ is cell velocity (eq. 3), $\vec{J}_{cond} = -D_{eff} \vec{\nabla} n_c$ is the conductive flux, $D_{eff}$ is the effective diffusion coefficient, $\vec{J}_i = k_i n_c \vec{\nabla} \phi_i$, are haptotaxis, electrotaxis, chemotaxis fluxes such that $\phi \equiv \rho$ is the matrix density for the haptotaxis, $\phi \equiv \phi_e$ is the electrostatic potential for the electrotaxis, $\phi \equiv c$ is the concentration of nutrients for the chemotaxis while $k_i$ are the model parameters which account for various types of interactions such as mechanical, electrostatic or chemical. Durotaxis is the phenomenon of directed CCM caused by the stiffness gradient of ECM. The durotaxis flux can be expressed as $\vec{J}_d = k_d n_c \Delta V_m (\vec{\nabla} E_m + \vec{\nabla} G_m)$ (where $k_d$ is the model parameter which represents a measure of matrix mobility induced by cell action, $\Delta V_m$ is the volume of a matrix part, $E_m$ is the matrix Young's modulus, and $G_m$ is the matrix shear modulus) (Pajic-Lijakovic and Milivojevic, 2021b). The Marangoni flux is a key factor which drives the cell sorting.

### 3.2 Total shear stress generation: constitutive models

The total SS generation caused by CCM is considered based on proposed constitutive viscoelastic models. Two models have been applied for describing viscoelasticity of cell monolayers, i.e. the Maxwell model and the Zener model (Lee and Wolgemuth, 2011; Notbohm et al., 2016; Pajic-Lijakovic and Milivojevic, 2019a;2020b). The Maxwell model, suitable for viscoelastic liquids, describes the stress relaxation under constant strain rate conditions, while the stain itself cannot relax (Pajic-Lijuakovic, 2021). This model has been proposed for a cell rearrangement in the form of weakly connected cell streams (Pajic-Lijakovic and Milivojevic, 2021a). Guevorkian et al. (2011) confirmed the Maxwell model as a suitable for describing the long-time cell rearrangement during the aggregate micropipette aspiration. The Zener model, suitable for viscoelastic solids, describes the stress relaxation under constant strain condition and the strain relaxation under constant stress condition (Pajic-Lijakovic, 2021). The stress relaxation corresponds to a time scale of minutes, while the strain changes caused by CCM and the residual stress accumulation correspond to a time scale of hours (Marmottant et al., 2009; Pajic-Lijakovic and Milivojevic, 2020c). Serra-Picamal et al. (2012) and Notbohm et al. (2016) measured the distribution of cell residual stress as function of strain caused by CCM within cell monolayers. They suggested the correlation of $\frac{d\tilde{\sigma}_{cR}}{d\tau} \sim \frac{d\tilde{\varepsilon}_c}{d\tau}$ (where $\tilde{\sigma}_{cR}$ is the cell residual stress, $\tilde{\varepsilon}_c$ is the corresponding cell strain, and $\tau$ is the long-time scale of hours). This result pointed to the Zener model as suitable for describing the viscoelasticity caused by CCM. In order to clarify this claim, we will discuss the Zener model and express the corresponding equation for the cell residual stress which arises from this model.

The Zener model is expressed as:

$$\tilde{\sigma}_{ci}(r,t,\tau) + \tau_{Ri} \dot{\tilde{\sigma}}_{ci} = G_i \tilde{\varepsilon}_{ci} + \eta_i \dot{\tilde{\varepsilon}}_{ci} \qquad (4)$$

where $i \equiv S, V$, $S$ is shear change, $V$ is volumetric change, $t$ is the time scale of minutes, $\tau$ is the time scale of hours, $\tilde{\sigma}_{ci}$ is the shear or normal stress, $\dot{\tilde{\sigma}}_{ci}$, is the rate of stress change, $\tilde{\varepsilon}_{ci}$ is the shear or volumetric strain ($\tilde{\varepsilon}_{cS} = \frac{1}{2}(\vec{\nabla}\vec{u} + \vec{\nabla}\vec{u}^T)$ is the shear strain, $\tilde{\varepsilon}_{cV} = \overline{(\vec{\nabla} \cdot \vec{u})}\tilde{I}$ is the volumetric strain, and $\vec{u}(r,\tau)$ is the local cell displacement field), $G_i$ is the shear or Young's elastic modulus, and $\eta_i$ is the shear or volumetric viscosity and $\tau_{Ri}$ is the stress relaxation time which corresponds to a time scale of minutes (Marmottant et al., 2009; Khalilgharibi et al., 2019). The stress relaxes during many short time relaxation cycles under constant strain per cycle, while strain change and the cell residual stress accumulation occur at a time scale of hours (Pajic-Lijakovic and Milivojevic, 2019a,2020c). Stress relaxation under constant strain $\tilde{\varepsilon}_{c0}(r,\tau)$ per single short time relaxation cycle is expressed starting from the initial condition $\tilde{\sigma}_i(r,t=0,\tau) = \tilde{\sigma}_0$ as:

$$\tilde{\sigma}_{ci}(r,t,\tau) = \tilde{\sigma}_0 e^{-\frac{t}{\tau_{Ri}}} + \tilde{\sigma}_{Ri}(r,\tau)^{ZM}\left(1 - e^{-\frac{t}{\tau_{Ri}}}\right) \qquad (5)$$

where $\tilde{\sigma}_{Ri}(r,\tau)^{ZM}$ is the residual shear or normal stress for the Zener model equal to $\tilde{\sigma}_{Ri}^{ZM} = G_i \tilde{\varepsilon}_{c0}$. This form of the cell residual stress ensures the correlations of $\tilde{\sigma}_{Ri}^{ZM} \sim \tilde{\varepsilon}_c$ as well as $\frac{d\tilde{\sigma}_{cR}}{d\tau} \sim \frac{d\tilde{\varepsilon}_c}{d\tau}$.

For further consideration, it is necessary to estimate the order of magnitude of the residual SS. The shear elastic modulus $G_S$ has been often calculated by using already measured value of the Young's modulus accompanied with the estimated value of the Poisson's ratio (Wu et al., 2018). The Poisson's ratio is in the range of $0.3 - 0.5$. Wu et al. (2018) reported that the modulus $G_S$ corresponds to a few kPa. The shear strain component per single stress relaxation cycle is: $\varepsilon_{cSxy} = \frac{\Delta u_x}{\Delta L}$ (where $\Delta u_x$ is cell displacement in x-direction and $\Delta L$ is the displacement gradient in y-direction). We supposed the shear strain component per single short-time relaxation cycle equal to $\varepsilon_{cSxy} \approx 0.01$. This value is proposed by introducing two assumptions: (1) $\Delta u_x \approx 1\ \mu m$ (which is an order of magnitude lower than the size of single-cells) and (2) $\Delta L \approx 100\ \mu m$ (which is an order of magnitude higher than the size of single cell). Clark and Vignjevic (2015) reported that the speed of migrating cell clusters during embryogenesis is about $v_c = 0.2 - 1\ \frac{\mu m}{min}$. The necessary time for cells to move the distance of $\Delta u_x \approx 1\ \mu m$ is equal to $\Delta t = \frac{\Delta u_x}{v_c}$ which is equal to $1 - 5\ min$. This calculated time corresponds to the stress relaxation time measured by Marmottant et al. (2009). Consequently, the calculated total cell residual SS corresponds to a few tens of Pa.

Notbohm et al. (2016) considered cell rearrangement within confluent monolayers and proposed the Maxwell model for characterizing the viscoelasticity caused by CCM. The Maxwell describes stress relaxation under constant strain rate and corresponding residual stress accumulation (Pajic-Lijakovic, 2021). Notbohm et al. (2016) measured the residual stress distribution within monolayers, while the stress relaxation wasn't considered. They pointed out that long-time change of cell residual stress correlates with the change of strain. However, this finding doesn't correspond to the Maxwell model. We will discuss the Maxwell model in order to extract the corresponding equation for the cell residual stress and clarify this issue. The Maxwell model is expressed as:

$$\tilde{\sigma}_i(r,t,\tau) + \tau_{Ri}\dot{\tilde{\sigma}}_i = \eta_i \dot{\tilde{\varepsilon}}_i(r,\tau) \qquad (6)$$

where $i \equiv S, V$, $S$ is shear change, $V$ is volumetric change. The stress relaxation under constant strain rate $\dot{\tilde{\varepsilon}}_{c0i}$ is expressed as:

$$\tilde{\sigma}_{ci}(r,t,\tau) = \tilde{\sigma}_0 e^{-\frac{t}{\tau_{Ri}}} + \tilde{\sigma}_{Ri}(r,\tau)^{MM}\left(1 - e^{-\frac{t}{\tau_{Ri}}}\right) \qquad (7)$$

where $\tilde{\sigma}_{Ri}(r,\tau)^{MM}$ is the residual shear or normal stress for the Maxwell model equal to $\tilde{\sigma}_{Ri}^{MM} = \eta_i \dot{\tilde{\varepsilon}}_{c0i}$. This formulation points to correlation between the residual stress and the strain rate ($\tilde{\sigma}_{Ri}^{MM} \sim \frac{d\tilde{\varepsilon}_c}{d\tau}$) rather than $\frac{d\tilde{\sigma}_{cR}}{d\tau} \sim \frac{d\tilde{\varepsilon}_c}{d\tau}$ which was experimentally confirmed by Notbohm et al. (2016). Pajic-Lijakovic and Milivojevic (2020a,b) supposed that 3D CCM induces significant energy dissipation within the biointerface between strongly connected migrating cell clusters and surrounding cells in the resting state. Consequently, they proposed the Maxwell model for the calculation of the cell residual SS within the biointerface. Pajic-Lijakovic and Milivojevic (2020a,b) expressed the residual SS within the biointerface as $\sigma_{c\,SR\,xz} \approx \eta \dot{\varepsilon}_{cS\,xz}$ (where $\dot{\varepsilon}_{cS\,xz} = \frac{\Delta v_c}{\Delta L}$ is the shear rate, $\Delta L$ is the biointerface thickness, and $\Delta v_c$ is the cell velocity difference across the biointerface). The calculated total residual SS is in the range of *15-73 Pa,* for the thickness of the biointerface equal to $\sim 100\ \mu m$, cell velocity in the range of $\sim 0.2 - 1\ \frac{\mu m}{min}$ (Clark and Vignjevic, 2015), and tissue viscosity equal to $\eta = 4.4x10^5\ Pas$ (Marmottant et al., 2009).

4. Cell swirling motion induced by the cell shear stress

While the SS caused by forced convection depends on the speed of migrating cells, the SS caused by the natural convection represents a consequence of established tissue surface tension gradient. Both contributions to the SS are the same order of magnitude, i.e. a several tens of Pa. The cell residual SS exerts work through the shear stress torque $\Delta\vec{T}(r,\tau) = \left(\vec{\nabla}\tilde{\sigma}_{cRS}(r,\tau)\right) X \vec{r}$ against the tissue cohesiveness quantified by the tissue surface tension $\gamma$ and can induce the cell swirling motion (Pajic-Lijakovic and Milivojevic, 2021b). The work of SS torque $\Delta\vec{T}$ is:

$$\Delta\vec{T} \cdot \vec{\omega}_c \geq \frac{d(\gamma(\vec{\nabla}\cdot\vec{n}))}{d\tau} \tag{8}$$

where $\vec{\omega}_c = \vec{\nabla} X \vec{v}_c$ is the angular velocity.

While the cell swirling motion has been observed during CCM of confluent cell monolayers (Notbohm et al., 2016; Chen et al., 2018), swirls formation has not appeared during a free expansion of monolayers under *in vitro* conditions (Serra-Picamal et al., 2012; Tambe et al., 2013). Lin et al. (2018) considered 2D CCM in the confined environment and pointed out that weak local cell polarity alignment (LA) and strong contact inhibition of locomotion (CIL) are the prerequisite for appearance of the cell swirling motion. These conditions correspond to a reduced tissue cohesiveness quantified by a lower tissue surface tension. Cell swirling motion induces a generation of mechanical standing waves (Notbohm et al., 2016; Pajic-Lijakovic and Milivojevic, 2020c). Cell within the swirl undergo successive radial extension and compression and azimuthal shear flow. Schematic presentation of the cell swirling motion is given on **Figure 2**.

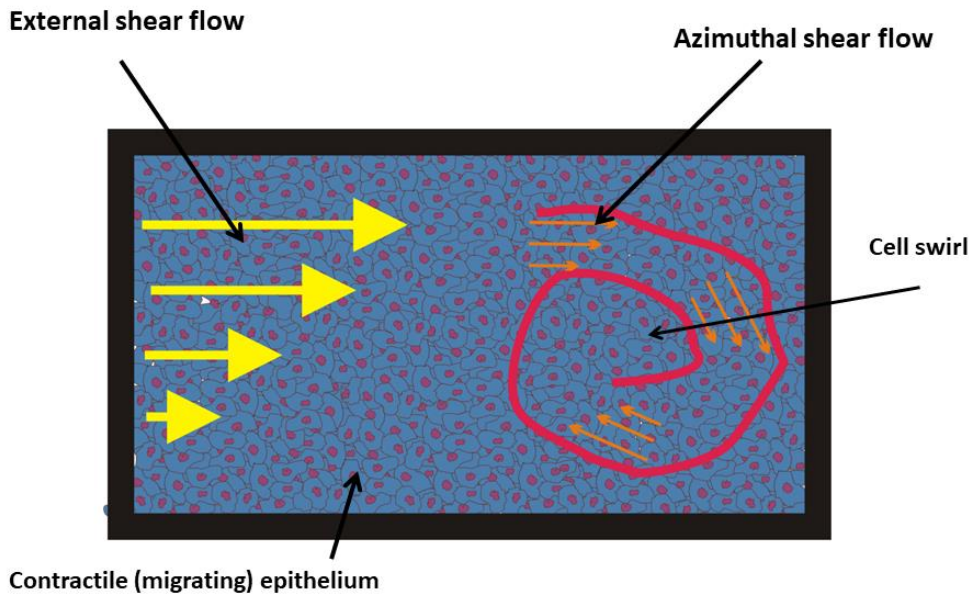

**Figure 2.** Shear flow induces the cell swirling motion.

The local azimuthal shear rate for cells within swirl could be lower than the shear rate out of swirl which induces the cell swirling motion, i.e. $\dot{\varepsilon}_{cS\,xy} > \dot{\varepsilon}_{cr\theta}{}^{SW}$ (where $\dot{\varepsilon}_{cS\,xy}$ is the shear rate capable of inducing the cell swirling motion equal to $\dot{\varepsilon}_{cS\,xy} = \frac{\Delta v_{cx}}{\Delta L}$, $\Delta v_{cx}$ is the difference in cell velocity in the x-direction, $\Delta L$ is the established velocity gradient in the y-direction, $\dot{\varepsilon}_{cr\theta}{}^{SW}$ is the azimuthal shear rate within a swirl, $\Delta v_{c\theta}$ is the azimuthal velocity difference, and $l_c$ is the size of single cells). Consequently, the cell swirling motion could represent a way of cells to avoid undesirable exposure to the SS. This important result should be tested experimentally. While the MDCK cells are capable of performing the cell swirling motion during 2D CCM (Notbohm et al., 2016), HaCaT skin cells and Caco2 intestinal cells forms an incomplete swirls under confluent condition (Peyret et al., 2019). It is in accordance with the fact that HaCaT skin cells and Caco2 intestinal cells establish stronger cell-cell adhesion contact in comparison with MDCK cells.

## 5. The Marangoni effect influences cell spreading and cell sorting

In the context of the Marangoni effect, we consider three types of mixed systems: (1) movement of the epithelial cell cluster through surrounding epithelium in the resting state, (2) movement of the mesenchymal cell cluster through surrounding epithelium in the (non-contractile) resting state, and (3) movement of the mesenchymal cell cluster through surrounding epithelium in the (contractile) migrating state. The Marangoni flux, as a consequence of the surface tension gradient between migrating cell cluster and surrounding cells, influences cell spreading from the regions of lower tissue surface tension to the regions of higher tissue surface tension and on that base influences the cell sorting. The phenomenon is schematically presented in **Figure 3**.

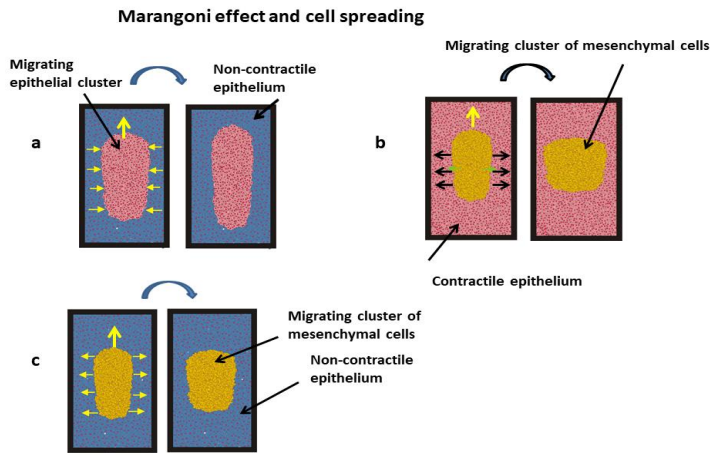

**Figure 3.** The Marangoni flux influences cell spreading (a) migrating epithelial cluster within non-contractile (resting) epithelium, (b) migrating cluster of mesenchymal cells within contractile epithelium, and (c) migrating cluster of mesenchymal cells within non-contractile epithelium.

The surface tensions of migrating cluster of mesenchymal cells $\gamma_{MC}^{c}$ satisfies the condition that: $\gamma_{MC}^{c} < \gamma_{EC}^{Nc} < \gamma_{EC}^{c}$ (where $\gamma_{MC}^{c}$ is the surface tension of contractile (migrating) mesenchymal cells, $\gamma_{EC}^{c}$ is the tissue surface tension of contractile epithelium, and $\gamma_{EC}^{Nc}$ is the tissue surface tension of non-contractile epithelium). Contractility of migrating mesenchymal cells intensifies the cell-cell repulsion which leads to a decrease in the tissue surface tension, such that $\gamma_{MC}^{c} < \gamma_{MC}^{Nc}$ (where $\gamma_{MC}^{Nc}$ is the tissue surface tension of the non-contractile mesenchymal cells) (Devenny et al., 2021). On the other hand, contractility of epithelial cells induces the reinforcement of E-cadherin mediated cell-cell adhesion contacts which leads to an increase in the tissue surface tension, such that: $\gamma_{EC}^{c} > \gamma_{EC}^{Nc}$ (Devenny et al., 2021).

Consequently, the spreading of cancer cells through active (contractile) epithelium is more efficient than their spreading through non-contractile epithelium caused by establishing the maximum surface tension gradient (Figure 3b,c). Wu et al. (2021) pointed out that in the environment of strong noise from migratory epithelial cells; tumor cells established coordinated movement while the movement of tumor cells through non-migratory epithelium was less persistent and more random. Lee et al. (2012) emphasized that the speed of single, mechanically soft breast carcinoma cells MDA-MB-231 is dramatically enhanced by surrounding stiff non-transformed MCF10A cells, compared with their movement through a monolayer of carcinoma cells. The MCF-10A breast epithelial cells form E-cadherin mediated AJs and fail to form TJs (Mohammed et al., 2021). The mesenchymal MDA-MB-231 cells are human breast cancer cell line obtained from a metastatic mammary adenocarcinoma (Devanny et al., 2021). While MCF10A cells in two-dimensional (2D) epithelial sheets exhibit aligned, directed motion and form compact spheroids in three-dimensional (3D) culture, the MDA-MB-231 cells in 2D epithelial sheets exhibit nonaligned, random motion and form invasive, noncontiguous clusters in 3D culture (Brückner et al., 2021). The MDA-MB-231 cells diffuse more persistently within a densely packed population than when they are free to crawl around with little interference (Lee and Lee, 2021). These results shed light on the role of stiff epithelial cells that neighbor individual cancer cells in early steps of cancer dissemination. Heine et al. (2021) found that the speed of MDA-MB-231 cells is 2-3 times larger than the speed of MCF10A within the same cell monolayer. The speed of MDA-MB-231 cells is significantly lower within the mixture monolayers made by two types of MDA-MB-231 cells (Heine et al., 2021). This result is expectable when we keep in mind that the mixed system of MDA-MB-231 - MCF10A cells establishes the maximum surface tension gradient. Devanny et al. (2021) considered 3D sorting of MDA-MB 436 – MCF-10A and MDA-MB 468 – MCF-10A multicellular systems. The sorting of the MDA-MB 436 – MCF-10A system was complete, while the MDA-MB 468 – MCF-10A established partial sorting. This result pointed to larger surface tension gradient established within MDA-MB 436 – MCF-10A system in comparison with MDA-MB 468 – MCF-10A system. The MCF-10A cells with higher tissue surface tension reached the core region, while the MDA-MB 436 covered the surface region of the multicellular spheroid.

**Conclusion**

Cells are very sensitive to the SS. The SS of a few *Pa* can induce severe damage to vimentin and keratin intermediate filament networks during 1 h, while shear stress of $\sim 60\ Pa$ influences gene expression which can cause the inflammation in epithelial cells during 5.5 h. The cell sensitivity to SS was discussed from the standpoint of rheology based on the mechanical properties of single semi-flexible filaments. Undesirable SS, generated during CCM, influences the biological processes such as: morphogenesis, wound healing and cancer invasion. It is necessary to indicate the causes of SS generation and estimate the order of magnitude of SS based on experimental data extracted from the literature and rheological modelling consideration. The SS generation corresponds to an order of magnitude of several tens of Pa and represents a consequence of natural and forced convection. While forced convection is dependent on cell speed, the natural convection is induced by the gradient of tissue surface tension (the Marangoni effect). The gradient of tissue surface tension guides the directed cell spreading from the regions of lower tissue surface tension to the regions of higher tissue surface tension. This directed cell migration is described by the Marangoni flux. The Marangoni flux is the the main factor responsible for cell sorting.

The phenomenon has been recognized during the rearrangement of (1) epithelial cell monolayers and (2) mixed cell monolayers made by epithelial and mesenchymal cells. However, a proper explanation has missed. Contractile (migrating) epithelial clusters have higher tissue surface tension than surrounding epithelium in the resting (non-contractile) state. It is in accordance with fact that the contractility of epithelial cells induces the reinforcement of E-cadherin mediated cell-cell adhesion contacts and on that base leads to an increase in the tissue surface tension. In contrast, the contractility of mesenchymal cells intensifies the repulsion between cells, which leads to a decrease in the tissue surface tension. Base of these findings, the important conclusion can be extracted. The Marangoni effect helps cancer cell spreading.

Additional experiments are necessary, in order to measure the surface tension gradient in various multicellular systems and estimate its influence on the directed cell migration which is of a crucial importance for various biological processes.

**Acknowledgment.** This work was supported by the Ministry of Education, Science and Technological Development of the Republic of Serbia (No. 451-03-68/2020-14/200135).

**Declaration of interest.** The author reports no conflict of interest.